\documentclass[prx, preprint, superscriptaddress, endfloats*]{revtex4-2}
\usepackage{amsfonts}
\usepackage{amsmath,hyperref,cleveref,graphicx,multirow,color,bm,textcomp,mathtools}

\usepackage{paralist}
\usepackage{srcltx}
\usepackage[all,cmtip]{xy}
\usepackage{mathrsfs}
\usepackage{srcltx,soul,subfigure}
\usepackage{threeparttable,dsfont,bbm}
\usepackage{color}
\usepackage{comment}
\usepackage{dcolumn}

\begin{document}
\title{Tunable High Spin Chern-Number Insulator Phases in Strained Sb Monolayer}
\author{Jacob Cook}
\affiliation {Department of Physics and Astronomy, University of Missouri, Columbia, Missouri 65211, USA}

\author{Po-Yuan Yang}
\affiliation {Department of Physics, National Cheng Kung University, Tainan 701, Taiwan}

\author{Theo Volz}
\affiliation {Rock Bridge High School, Columbia, Missouri 65203, USA}

\author{Clayton Conner}
\affiliation {Department of Physics and Astronomy, University of Missouri, Columbia, Missouri 65211, USA}

\author{Riley Satterfield}
\affiliation {Department of Physics and Astronomy, University of Missouri, Columbia, Missouri 65211, USA}

\author{Joseph Berglund}
\affiliation {Department of Physics and Astronomy, University of Missouri, Columbia, Missouri 65211, USA}

\author{Qiangsheng Lu}
\affiliation {Department of Physics and Astronomy, University of Missouri, Columbia, Missouri 65211, USA}
\affiliation {Materials Science and Technology Division, Oak Ridge National Laboratory, Oak Ridge, Tennessee 37831 USA}

\author{Rob~G.~Moore}
\affiliation {Materials Science and Technology Division, Oak Ridge National Laboratory, Oak Ridge, Tennessee 37831 USA}

\author{Yueh-Ting Yao}
\affiliation {Department of Physics, National Cheng Kung University, Tainan 701, Taiwan}

\author{Tay-Rong Chang*}
\affiliation {Department of Physics, National Cheng Kung University, Tainan 701, Taiwan}
\affiliation {Center for Quantum Frontiers of Research and Technology (QFort), Tainan, 70101, Taiwan}
\affiliation {Physics Division, National Center for Theoretical Sciences, Taipei, 10617, Taiwan}

\author{Guang~Bian*}\email{email: u32trc00@phys.ncku.edu.tw, biang@missouri.edu}
\affiliation {Department of Physics and Astronomy, University of Missouri, Columbia, Missouri 65211, USA}
\affiliation {MU Materials Science \& Engineering Institute, University of Missouri, Columbia, Missouri 65211, USA}

\newpage
\begin{abstract}
{High spin Chern-number insulators (HSCI) have emerged as a novel 2D topological phase of condensed matter that is beyond the classification of topological quantum chemistry. In this work, we report the observation of a semimetallic Sb monolayer carrying the same band topology as HSCI with a spin Chern number equal to 2. Our calculations further indicate a moderate lattice strain can make Sb monolayer an insulator or a semimetal with a tunable spin Chern number from 0 to 3. The results suggest strained Sb monolayers as a promising platform for exploring exotic properties of the HSCI topological matter.}

\end{abstract}


\maketitle

\newpage
Topological phases of condensed matter are of significant interest for their novel electronic properties and device applications \cite{Hasan_2010,Qi_2011,Bansil_2016,Xiao_2021}. Most topological materials can be classified in the framework of topological quantum chemistry (TQC) using symmetry indicators (SI) at high-symmetry points (HSP)\cite{Slager2013, Bradlyn_2017, PhysRevX.7.041069, Tang_2019}. Typically, band inversions occur at HSPs and play a crucial role in defining the non-trivial bulk topology. Recently, high spin Chern-number insulator (HSCI) has emerged as a novel topological phase that is beyond the description of TQC\cite{Ezawa2013, Wang_2022, Bai_2022, PhysRevResearch.5.033013, Peng_2023}. In HSCIs, band inversion occurs at an even number of non-HSP momenta, leading to a zero $\mathbb{Z}_2$ topological invariant. Thus, this 2D system with time-reversal symmetry is regarded as topologically trivial according to the TQC classification. On the other hand, it has theoretically demonstrated that in the $\alpha$ phase of Sb and Bi monolayers \cite{Wang_2022, Bai_2022}, a nontrivial distribution of Berry curvature around the momenta where band inversion occurs gives rise to a high spin Chern number $\mathcal{C}_S$=2. Consequently, the HSCI system harbors two pairs of gapless helical edge states, which can drive a larger spin-polarized current. It has also been proven that this new type of bulk-boundary correspondence in HSCI is robust even in the presence of spin-orbit coupling (SOC) due to the protection of a ``hidden'' feature spectrum topology \cite{feature, Chang_2024}. Therefore, there is a pressing need to realize an experimentally viable HSCI to facilitate the exploration of the exotic behaviors associated with this unconventional topological matter.

In this work, we synthesized $\alpha$-Sb monolayer on SnSe substrate by molecular beam epitaxy (MBE). Our angle-resolved photoemission (ARPES) measurements and first-principles calculations indicate that the nearly freestanding $\alpha$-Sb monolayer on SnSe, though semimetallic without a global band gap, shares the same band topology as HSCI with spin Chern number equal to 2.  Our calculations further show that applying a moderate uniaxial lattice strain ($<$2\% in the zigzag direction) makes $\alpha$-Sb an insulator while retaining the nontrivial HSCI band topology. Interestingly, Interestingly, increasing the lattice strain can induce an additional band inversion at the time-reversal invariant momentum point $\Gamma$. Consequently, the system transforms into a topological semimetallic phase characterized by a nonzero $\mathbb{Z}_2$ number. The spin Chern number also increases by 1 in this process. Therefore, rich topological phases with a tunable spin Chern number ranging from 0 to 3 can be achieved in the $\alpha$-Sb monolayer system by applying lattice strains. The proposed lattice strains can be readily achieved by selecting suitable substrates for MBE synthesis, as ultrathin epitaxial films are particularly sensitive to substrate conditions. A comprehensive phase diagram of $\alpha$-Sb under various strains is presented, serving as a valuable guide for future experimental exploration of the exotic HSCI phase with tailored electronic and spintronic properties. 

Sb monolayer in $\alpha$ phase has a black phosphorous (BP) crystal structure with the $Pnma$ space group as shown Figs.~1(a-c) \cite{Sb_Li_2020, Märkl_2018, Bi_Meta_2009}. The two Sb atomic sublayers (plotted in different colors) form a buckled structure. $\alpha$-Sb monolayer was grown by MBE on a cleaved (001)-surface of SnSe. The STM image (Fig.~1(d)) demonstrates a good uniformity of the MBE sample. The height profile taken along the green arrow (marked in Fig. 1(d)) shows that the apparent height of the epitaxial monolayer on the SnSe surface is 6.1$\pm 0.1 \mathrm{\r{A}}$. The surface unit cell of the Sb monolayer can be seen in the zoom-in STM image in Fig.~1(e). The averaged in-plane lattice constants of 1L $\alpha$-Sb are extracted from the height profiles shown in Figs.~1(f,g), that is  $a$ = 4.29 $\pm$ 0.1 $\mathrm{\r{A}}$ and $b$ = 4.77 $\pm$ 0.1 $\mathrm{\r{A}}$ in the $x$ and $y$-directions, respectively. 

 The band structure of the MBE-grown 1ML $\alpha$-Sb was mapped out by ARPES (Figs.~1(h-j)). The measured  (Fig.~1(j)) shows three prominent features near the Fermi level, namely, an electron pocket at $\Gamma$ from the conduction band and two hole pockets centered at $(k_{x}, k_{y})=(0, \pm 0.49~\mathrm{\AA}^{-1})$ from the valence band. The spectrum along the high-symmetry direction $\Gamma-$Y is plotted in Fig.~1(h). The overlaid white solid curves are the first-principles band dispersion based on experimental lattice parameters. Due to the presence of substrate, the measured band structure has a shift of ~+0.1 eV compared to the DFT result. The good agreement between the ARPES spectrum and the calculated bands indicates the system is semimetallic with a small negative gap between the conduction and valence bands. The ARPES spectral cut taken along the line ``Cut2'' at $k_{y}=0.49\ \mathrm{\AA}^{-1}$ perpendicular to $\Gamma-$Y (Fig.~1(i)) shows the hole-like valence band barely touches the Fermi level and is separated from the conduction band by a small energy gap of 0.15~eV. The band dispersion resembles a gaped conical surface. The conical hole pockets at $(k_{x}, k_{y})=(0, \pm 0.49\ \mathrm{\AA}^{-1})$ are referred to as two valleys in the following discussions.  
 
Now we discuss the band topology of 1ML $\alpha$-Sb. The high spin Chern number can be attributed to the band inversion at the two valleys of 1ML $\alpha$-Sb.  A schematic of band inversion at the valleys is shown in Figs.~2(a-d). Starting from a conventional semiconducting band structure in the absence of SOC (Fig.~2(a)). The band gap can be completely closed by continuously tuning a parameter of the system such as lattice constants, resulting in a 2D Dirac nodal point (Fig.~2(b)). Further tuning the parameter causes a band inversion between the conduction and valence bands and generates two Dirac points (DPs) marked by red dots in Fig.~2(c). (We note that the DPs generated this way are located at generic momenta rather than HSPs. Those DPs are referred to in the literature as ``unpinned'' Dirac states \cite{Lu2016}. The inclusion of the spin-orbit coupling gaps out the unpinned Dirac states (Fig.~2(d)) and induces a nontrivial Berry curvature around the DPs \cite{Lu_2022, Rashba_2013, PhysRevB.88.085427}. This process of band inversion can be seen in the band dispersion of 1ML~$\alpha$-Sb with varied lattice parameters around the valley momenta (0,$\pm\Lambda_y$). Figures~2(e-g) show the calculated band structure of 1ML~$\alpha$-Sb with lattice constants $a=4.35\mathrm{\AA}$, $b=5.11, 4.98, 4.62\mathrm{\AA}$, respectively. The bands near the valley momentum $\Lambda_y$ primarily originate from the $p_{x,y}$ orbitals of Sb atoms. The energy gap at the valley closes and reopens as the lattice parameter $b$ decreases from 5.11\AA\ to 4.62\AA. The gap vanishes at the critical lattice values of (4.35\AA, 4.98\AA). Figures~2(h-j) depict the valley band dispersion calculated along the line perpendicular to $\Gamma-\mathrm{Y}$ with lattice parameters corresponding to those in Figs.~2(e-g). The band dispersion in Fig.~2(j) demonstrates the SOC-induced band gap (similar to that in Fig~2(d)) and an inverted order between the conduction and valence bands at the valley momentum (0, $\Lambda_y$). The band inversion at the two valleys is marked by the red dots and cycles in the 1st Brillouin zone of 1ML~$\alpha$-Sb in Fig.~2(l). The contributions to spin Chern number from the Berry curvature distributions at the two valleys add up and result in a high spin Chern number $\mathcal{C}_S$=2. The even spin Chern number indicates that there exist two pairs of gapless spin-polarized edge states. This new type of bulk-boundary correspondence is guaranteed by a ``hidden'' feature spectrum topology as reported in the previous work\cite{feature}. Meanwhile, the double band inversion at generic momenta (0,$\pm\Lambda_y$) gives a trivial $\mathbb{Z}_2$ topological invariant, thus indicating the 1ML~$\alpha$-Sb with even high spin Chern numbers is a new type of topological materials beyond the classification of TQC. Moreover, an extra band inversion can be induced at the time-reversal invariant momentum (TRIM) $\Gamma$ by tuning the lattice parameter as shown in Fig.~2(k). The band inversion occurs between the conduction and valence bands of $p_{z}$ orbital character. A tiny SOC-induced gap of ~15 meV is opened at band crossing points (see Supplementary Materials). This band inversion at $\Gamma$ gives rise to a nonzero $\mathbb{Z}_2$ topological invariant and makes the system a topological insulator. The contribution to spin Chern number from this band inversion has the same sign as those from two valley band inversions. Therefore, a topological phase with spin Chern number equal to 3 can be achieved in $\alpha$-Sb monolayer by tuning the lattice strains to create three band inversions (one at $\Gamma$ and two at valleys) simultaneously.

To demonstrate the nontrivial band topology and protected helical edge states, we calculated the edge state bands and spin-resolved Wilson loops of 1ML~$\alpha$-Sb with different lattice constants. The results are summarized in Fig.~3. Here we focus on four topologically distinct phases. Figure~3(a) shows the band structure of 1ML~$\alpha$-Sb with $a=4.16\mathrm{\AA}$ and $b=4.57\mathrm{\AA}$, in which we found band inversions at $\Gamma$ and two valleys. Figure~3(f) plots the projected edge spectrum of a semi-infinite 1ML~$\alpha$-Sb with an open edge along the armchair ($\overline{\Gamma}-\overline{\mathrm{Y}}$) direction. The solid curves within the projected bulk band gap depict the dispersion of edge state bands. The spin polarization of the edge state bands is shown in Fig.~3(k). There exist three pairs of spin-polarized edge state bands traversing the bulk band gap between the projected conduction and valence bands, indicating the spin Chern number is equal to 3. This property is further confirmed by the calculation of the spin-resolved Wilson loop shown in Fig.~3(p). The red (blue) curves demonstrate the evolution of the Wannier center $\theta$ of spin-up (spin-down) valence bands as $k_y$ sweeps from one end to the other end of the Brillouin zone. The Wannier center of each spin branch traverses the unit cell three times in the winding of the Wilson loop, proving that the spin Chern number is 3 \cite{Chang_2024}. In this case, the conduction and valence bands overlap in energy, and the $\mathbb{Z}_2$ topological invariant is $v=1$ due to the band inversion at $\Gamma$. Therefore, this phase is a topological semimetal with an odd high spin Chern number (HSC-TSM for short). Changing $(a, b)$ from ($4.16\mathrm{\AA}, 4.57\mathrm{\AA}$) to ($4.42\mathrm{\AA}, 4.77\mathrm{\AA}$) restores the normal band order at $\Gamma$, retains the valley band inversions, and generates an absolute band gap as shown in Fig.~3(b). The calculated edge state spectrum and spin-resolved Wilson loop (Figs.~3(g,l,q)) demonstrate this phase is a high spin Chern-number insulator (HSCI) with $\mathcal{C}_S$=2. The band structure, edge state spectrum, and Wilson loop based on our experimental lattice constants ($a = 4.29\mathrm{\AA}$, $b = 4.77\mathrm{\AA}$) are plotted in Figs.~3(c,h,m,r). Our experimental $\alpha$-Sb monolayer shares the same band topology and spin Chern number as the HSCI in Fig.~3(b) but it is with a negative band gap. So the $\alpha$-Sb monolayer grown in our experiment is a high spin Chern-number semimetal (HSC-SM). At $(a, b)$=$(4.08\mathrm{\AA}, 4.77\mathrm{\AA})$, a band inversion occurs only at the TRIM point $\Gamma$, leading to a nonzero $\mathbb{Z}_2$ topological invariant and an odd spin Chern number, $\mathcal{C}_S$=1. So the system is a topological semimetal with a negative band gap as shown in Figs.~3(d,i,n,s). By contrast, $\alpha$-Sb monolayer with $(a, b)$=$(4.16\mathrm{\AA}, 4.96\mathrm{\AA})$ exhibits no band inversion in the entire Brillouin zone and thus becomes a topologically trivial semimetal (Figs.~3(e,j,o,t)) with zero $\mathbb{Z}_2$ and $\mathcal{C}_S$ numbers.

 Topologically distinct phases of $\alpha$-Sb monolayer can be achieved by applying moderate lattice strains. To understand the influence of lattice strains on the band topology of $\alpha$-Sb monolayer, we performed systematic simulations of the band structure with tensile (positive, up to $5\%$) and compressive (negative, up to $-5\%$) strains (based on our experimental lattice values). The results are shown in Fig.~4. In particular, we extracted the absolute values of the global band gap ($\Delta \mathrm{E_A}$), the band gap at the valleys ($\Delta \mathrm{E_V}$), and the band gap at $\Gamma$ ($\Delta \mathrm{E_{\Gamma}}$), as defined in Fig.~4(a). $\Delta \mathrm{E_V}$ and $\Delta \mathrm{E_{\Gamma}}$ signify the band inversions at two valleys and $\Gamma$, respectively. $\Delta \mathrm{E_A}$ determines the transport properties of the system. A positive $\Delta \mathrm{E_A}$ indicates the system is an insulator/semiconductor, whereas a zero or negative $\mathrm{E_A}$ corresponds to a semimetallic phase. A plateau of quantized spin Hall conductance is expected to appear within the bulk band gap in the HSCI phase with $\Delta \mathrm{E_A} >0$. To show this feature, we calculated the spin Hall conductance of $\alpha$-Sb monolayer with nonzero spin Chern numbers. The results are plotted in Figs.~4(b,c). The $\alpha$-Sb monolayer grown in our experiment is a high spin Chern-number semimetal ($\mathcal{C}_S=2$) without a global band gap. The spin Hall conductance is nonzero near the Fermi level (Fig.~4(b)), but not quantized due to the contribution from the bulk band pockets at the Fermi level. By contrast,  a plateau of value $2\times\frac{e^2}{h}$ shows up in the calculated spin Hall conductance curve of $\alpha$-Sb monolayer  (with $(a, b) = (4.42\mathrm{\AA}, 4.77\mathrm{\AA})$) in the HSCI phase (Fig.~4(c)). The quantized spin Hall conductance can be attributed to the fact that the SOC terms that violate the conservation of $s_z$ in the effective Hamiltonian are highly suppressed \cite{Ahn_2017}. Specifically, only the SOC coupling arising from the interlayer hopping between the two sublayers (the ``Upper" and ``Lower" Sb layers in Fig. 1(b)) can break the conservation of $s_z$. On the other hand, the bands near the valleys predominately originate from in-plane $p_{x,y}$ orbitals. So, the $s_z$-violating terms from the interlayer hopping are almost vanishing. Consequently, $s_z$ can be considered as a good quantum number in this system, leading to the quantized spin-Hall conductance in the HSCI phase.
 
  The maps of $\Delta \mathrm{E_V}$, $\Delta \mathrm{E_\Gamma}$, and $\Delta \mathrm{E_A}$ under different strains in {\it x} and {\it y} directions are plotted in Fig.~4(d), respectively. The red dot marks our experimental lattice constants of $\alpha$-Sb monolayer. The red solid lines (with $\Delta \mathrm{E_{V}}=0$ or $\Delta \mathrm{E_{\Gamma}}=0$ depict the boundary between phases with different spin Chern numbers. Combining the results from Fig.~4(d), we can get the phase diagram of $\alpha$-Sb monolayer under strains (Fig.~4(e)), which allows straightforward identification of the band properties of the strained $\alpha$-Sb films. For example, our MBE-grown sample (labeled with ``Exp") lies in the region of ``high spin Chern-number semimetal'', and a moderate tensile strain of 2\% in {\it x} direction can make it an HSCI. The lattice parameter values of $\alpha$-Sb monolayer extracted from previous works \cite{Shi_2020, Salehitaleghani_2023, Märkl_2018, Shi_2019, Li_2022, Bai_2022, Niu_2021, Lu_2021} are marked in the phase diagram. Remarkably, six topologically distinct phases with the spin Chern number ranging from 0 to 3 can be readily achieved in $\alpha$-Sb monolayer by moderate lattice strains. The lattice constants of epitaxial monolayers can be effectively controlled, as the lattice is highly susceptible to substrate effects\cite{Datye_2022, PhysRevB.80.245407}, such as charge transfer\cite{Shi_2020}, interlayer bonding\cite{Zhang_2021}, and interfacial strains\cite{Liu_2022,Li_2022,Cook_2023,Halbertal_2023}. Tensile and compressive lattice strains of up to 5\% have been reported in 2D van der Waals (vdW) materials \cite{DENG201814}. The robustness of vdW materials to lattice distortions makes them ideal for strain engineering of electronic structures.

In conclusion, our ARPES and first-principles results demonstrated that the epitaxial $\alpha$-Sb monolayer grown on SnSe substrate is a high spin Chern-number semimetal with $\mathcal{C}_S$=2. Rich topological phases with tunable spin Chern numbers can be realized by applying moderate lattice stains. Our work sheds light on materials engineering of the novel HSCI topological phases in epitaxial atomic layers and offers a promising avenue for the development of novel electronic and spintronic devices based on the multiple spin-polarized edge channels of HSCI.

\section*{acknowledgement}

J.C. and P.-Y.Y. contributed equally to this work. The work at the University of Missouri was supported by the U.S. Department of Energy, Office of Basic Energy Sciences, Division of Materials Science and Engineering, under Grant No.~DE-SC0024294. G.B. was supported by the Gordon and Betty Moore Foundation, grant DOI 10.37807/gbmf12247. Portions of this work at Oak Ridge National Laboratory were supported by the U.S. Department of Energy, Office of Science, National Quantum Information Science Research Centers, Quantum Science Center (Q. L. and R. G. M.). T.-R.C. was supported by the 2030 Cross-Generation Young Scholars Program from the National Science and Technology Council (NSTC) in Taiwan (Program No. MOST111-2628-M-006-003-MY3), National Cheng Kung University (NCKU), Taiwan, and National Center for Theoretical Sciences, Taiwan. This research was supported, in part, by Higher Education Sprout Project, Ministry of Education to the Headquarters of University Advancement at NCKU.

\bibliography{Sb_HSCI}

\begin{thebibliography}{36}%
\makeatletter
\providecommand \@ifxundefined [1]{%
 \@ifx{#1\undefined}
}%
\providecommand \@ifnum [1]{%
 \ifnum #1\expandafter \@firstoftwo
 \else \expandafter \@secondoftwo
 \fi
}%
\providecommand \@ifx [1]{%
 \ifx #1\expandafter \@firstoftwo
 \else \expandafter \@secondoftwo
 \fi
}%
\providecommand \natexlab [1]{#1}%
\providecommand \enquote  [1]{``#1''}%
\providecommand \bibnamefont  [1]{#1}%
\providecommand \bibfnamefont [1]{#1}%
\providecommand \citenamefont [1]{#1}%
\providecommand \href@noop [0]{\@secondoftwo}%
\providecommand \href [0]{\begingroup \@sanitize@url \@href}%
\providecommand \@href[1]{\@@startlink{#1}\@@href}%
\providecommand \@@href[1]{\endgroup#1\@@endlink}%
\providecommand \@sanitize@url [0]{\catcode `\\12\catcode `\$12\catcode `\&12\catcode `\#12\catcode `\^12\catcode `\_12\catcode `\%12\relax}%
\providecommand \@@startlink[1]{}%
\providecommand \@@endlink[0]{}%
\providecommand \url  [0]{\begingroup\@sanitize@url \@url }%
\providecommand \@url [1]{\endgroup\@href {#1}{\urlprefix }}%
\providecommand \urlprefix  [0]{URL }%
\providecommand \Eprint [0]{\href }%
\providecommand \doibase [0]{https://doi.org/}%
\providecommand \selectlanguage [0]{\@gobble}%
\providecommand \bibinfo  [0]{\@secondoftwo}%
\providecommand \bibfield  [0]{\@secondoftwo}%
\providecommand \translation [1]{[#1]}%
\providecommand \BibitemOpen [0]{}%
\providecommand \bibitemStop [0]{}%
\providecommand \bibitemNoStop [0]{.\EOS\space}%
\providecommand \EOS [0]{\spacefactor3000\relax}%
\providecommand \BibitemShut  [1]{\csname bibitem#1\endcsname}%
\let\auto@bib@innerbib\@empty
\bibitem [{\citenamefont {Hasan}\ and\ \citenamefont {Kane}(2010)}]{Hasan_2010}%
  \BibitemOpen
  \bibfield  {author} {\bibinfo {author} {\bibfnamefont {M.~Z.}\ \bibnamefont {Hasan}}\ and\ \bibinfo {author} {\bibfnamefont {C.~L.}\ \bibnamefont {Kane}},\ }\bibfield  {title} {\bibinfo {title} {Colloquium: Topological insulators},\ }\href {https://doi.org/10.1103/RevModPhys.82.3045} {\bibfield  {journal} {\bibinfo  {journal} {Rev. Mod. Phys.}\ }\textbf {\bibinfo {volume} {82}},\ \bibinfo {pages} {3045} (\bibinfo {year} {2010})}\BibitemShut {NoStop}%
\bibitem [{\citenamefont {Qi}\ and\ \citenamefont {Zhang}(2011)}]{Qi_2011}%
  \BibitemOpen
  \bibfield  {author} {\bibinfo {author} {\bibfnamefont {X.-L.}\ \bibnamefont {Qi}}\ and\ \bibinfo {author} {\bibfnamefont {S.-C.}\ \bibnamefont {Zhang}},\ }\bibfield  {title} {\bibinfo {title} {Topological insulators and superconductors},\ }\href {https://doi.org/10.1103/RevModPhys.83.1057} {\bibfield  {journal} {\bibinfo  {journal} {Rev. Mod. Phys.}\ }\textbf {\bibinfo {volume} {83}},\ \bibinfo {pages} {1057} (\bibinfo {year} {2011})}\BibitemShut {NoStop}%
\bibitem [{\citenamefont {Bansil}\ \emph {et~al.}(2016)\citenamefont {Bansil}, \citenamefont {Lin},\ and\ \citenamefont {Das}}]{Bansil_2016}%
  \BibitemOpen
  \bibfield  {author} {\bibinfo {author} {\bibfnamefont {A.}~\bibnamefont {Bansil}}, \bibinfo {author} {\bibfnamefont {H.}~\bibnamefont {Lin}},\ and\ \bibinfo {author} {\bibfnamefont {T.}~\bibnamefont {Das}},\ }\bibfield  {title} {\bibinfo {title} {Colloquium: Topological band theory},\ }\href {https://doi.org/10.1103/RevModPhys.88.021004} {\bibfield  {journal} {\bibinfo  {journal} {Rev. Mod. Phys.}\ }\textbf {\bibinfo {volume} {88}},\ \bibinfo {pages} {021004} (\bibinfo {year} {2016})}\BibitemShut {NoStop}%
\bibitem [{\citenamefont {Xiao}\ and\ \citenamefont {Yan}(2021)}]{Xiao_2021}%
  \BibitemOpen
  \bibfield  {author} {\bibinfo {author} {\bibfnamefont {J.}~\bibnamefont {Xiao}}\ and\ \bibinfo {author} {\bibfnamefont {B.}~\bibnamefont {Yan}},\ }\bibfield  {title} {\bibinfo {title} {First-principles calculations for topological quantum materials},\ }\href {https://doi.org/10.1038/s42254-021-00292-8} {\bibfield  {journal} {\bibinfo  {journal} {Nature Reviews Physics}\ }\textbf {\bibinfo {volume} {3}},\ \bibinfo {pages} {283} (\bibinfo {year} {2021})}\BibitemShut {NoStop}%
\bibitem [{\citenamefont {Slager}\ \emph {et~al.}(2013)\citenamefont {Slager}, \citenamefont {Mesaros}, \citenamefont {Juri{\v{c}}i{\'{c}}},\ and\ \citenamefont {Zaanen}}]{Slager2013}%
  \BibitemOpen
  \bibfield  {author} {\bibinfo {author} {\bibfnamefont {R.-J.}\ \bibnamefont {Slager}}, \bibinfo {author} {\bibfnamefont {A.}~\bibnamefont {Mesaros}}, \bibinfo {author} {\bibfnamefont {V.}~\bibnamefont {Juri{\v{c}}i{\'{c}}}},\ and\ \bibinfo {author} {\bibfnamefont {J.}~\bibnamefont {Zaanen}},\ }\bibfield  {title} {\bibinfo {title} {The space group classification of topological band-insulators},\ }\href {https://doi.org/10.1038/nphys2513} {\bibfield  {journal} {\bibinfo  {journal} {Nature Physics}\ }\textbf {\bibinfo {volume} {9}},\ \bibinfo {pages} {98} (\bibinfo {year} {2013})}\BibitemShut {NoStop}%
\bibitem [{\citenamefont {Bradlyn}\ \emph {et~al.}(2017)\citenamefont {Bradlyn}, \citenamefont {Elcoro}, \citenamefont {Cano}, \citenamefont {Vergniory}, \citenamefont {Wang}, \citenamefont {Felser}, \citenamefont {Aroyo},\ and\ \citenamefont {Bernevig}}]{Bradlyn_2017}%
  \BibitemOpen
  \bibfield  {author} {\bibinfo {author} {\bibfnamefont {B.}~\bibnamefont {Bradlyn}}, \bibinfo {author} {\bibfnamefont {L.}~\bibnamefont {Elcoro}}, \bibinfo {author} {\bibfnamefont {J.}~\bibnamefont {Cano}}, \bibinfo {author} {\bibfnamefont {M.~G.}\ \bibnamefont {Vergniory}}, \bibinfo {author} {\bibfnamefont {Z.}~\bibnamefont {Wang}}, \bibinfo {author} {\bibfnamefont {C.}~\bibnamefont {Felser}}, \bibinfo {author} {\bibfnamefont {M.~I.}\ \bibnamefont {Aroyo}},\ and\ \bibinfo {author} {\bibfnamefont {B.~A.}\ \bibnamefont {Bernevig}},\ }\bibfield  {title} {\bibinfo {title} {Topological quantum chemistry},\ }\href {https://doi.org/10.1038/nature23268} {\bibfield  {journal} {\bibinfo  {journal} {Nature}\ }\textbf {\bibinfo {volume} {547}},\ \bibinfo {pages} {298} (\bibinfo {year} {2017})}\BibitemShut {NoStop}%
\bibitem [{\citenamefont {Kruthoff}\ \emph {et~al.}(2017)\citenamefont {Kruthoff}, \citenamefont {de~Boer}, \citenamefont {van Wezel}, \citenamefont {Kane},\ and\ \citenamefont {Slager}}]{PhysRevX.7.041069}%
  \BibitemOpen
  \bibfield  {author} {\bibinfo {author} {\bibfnamefont {J.}~\bibnamefont {Kruthoff}}, \bibinfo {author} {\bibfnamefont {J.}~\bibnamefont {de~Boer}}, \bibinfo {author} {\bibfnamefont {J.}~\bibnamefont {van Wezel}}, \bibinfo {author} {\bibfnamefont {C.~L.}\ \bibnamefont {Kane}},\ and\ \bibinfo {author} {\bibfnamefont {R.-J.}\ \bibnamefont {Slager}},\ }\bibfield  {title} {\bibinfo {title} {Topological classification of crystalline insulators through band structure combinatorics},\ }\href {https://doi.org/10.1103/PhysRevX.7.041069} {\bibfield  {journal} {\bibinfo  {journal} {Phys. Rev. X}\ }\textbf {\bibinfo {volume} {7}},\ \bibinfo {pages} {041069} (\bibinfo {year} {2017})}\BibitemShut {NoStop}%
\bibitem [{\citenamefont {Tang}\ \emph {et~al.}(2019)\citenamefont {Tang}, \citenamefont {Po}, \citenamefont {Vishwanath},\ and\ \citenamefont {Wan}}]{Tang_2019}%
  \BibitemOpen
  \bibfield  {author} {\bibinfo {author} {\bibfnamefont {F.}~\bibnamefont {Tang}}, \bibinfo {author} {\bibfnamefont {H.~C.}\ \bibnamefont {Po}}, \bibinfo {author} {\bibfnamefont {A.}~\bibnamefont {Vishwanath}},\ and\ \bibinfo {author} {\bibfnamefont {X.}~\bibnamefont {Wan}},\ }\bibfield  {title} {\bibinfo {title} {Comprehensive search for topological materials using symmetry indicators},\ }\href {https://doi.org/10.1038/s41586-019-0937-5} {\bibfield  {journal} {\bibinfo  {journal} {Nature}\ }\textbf {\bibinfo {volume} {566}},\ \bibinfo {pages} {486} (\bibinfo {year} {2019})}\BibitemShut {NoStop}%
\bibitem [{\citenamefont {Ezawa}(2013)}]{Ezawa2013}%
  \BibitemOpen
  \bibfield  {author} {\bibinfo {author} {\bibfnamefont {M.}~\bibnamefont {Ezawa}},\ }\bibfield  {title} {\bibinfo {title} {High spin-chern insulators with magnetic order},\ }\href {https://doi.org/10.1038/srep03435} {\bibfield  {journal} {\bibinfo  {journal} {Scientific Reports}\ }\textbf {\bibinfo {volume} {3}},\ \bibinfo {pages} {3435} (\bibinfo {year} {2013})}\BibitemShut {NoStop}%
\bibitem [{\citenamefont {Wang}\ \emph {et~al.}(2022)\citenamefont {Wang}, \citenamefont {Zhou}, \citenamefont {Lin}, \citenamefont {Lin},\ and\ \citenamefont {Bansil}}]{Wang_2022}%
  \BibitemOpen
  \bibfield  {author} {\bibinfo {author} {\bibfnamefont {B.}~\bibnamefont {Wang}}, \bibinfo {author} {\bibfnamefont {X.}~\bibnamefont {Zhou}}, \bibinfo {author} {\bibfnamefont {Y.-C.}\ \bibnamefont {Lin}}, \bibinfo {author} {\bibfnamefont {H.}~\bibnamefont {Lin}},\ and\ \bibinfo {author} {\bibfnamefont {A.}~\bibnamefont {Bansil}},\ }\href@noop {} {\bibinfo {title} {High spin-chern-number insulator in $\alpha$-antimonene with a hidden topological phase}} (\bibinfo {year} {2022}),\ \Eprint {https://arxiv.org/abs/2202.04162} {arXiv:2202.04162 [cond-mat.mes-hall]} \BibitemShut {NoStop}%
\bibitem [{\citenamefont {Bai}\ \emph {et~al.}(2022)\citenamefont {Bai}, \citenamefont {Cai}, \citenamefont {Mao}, \citenamefont {Li}, \citenamefont {Dai}, \citenamefont {Huang},\ and\ \citenamefont {Niu}}]{Bai_2022}%
  \BibitemOpen
  \bibfield  {author} {\bibinfo {author} {\bibfnamefont {Y.}~\bibnamefont {Bai}}, \bibinfo {author} {\bibfnamefont {L.}~\bibnamefont {Cai}}, \bibinfo {author} {\bibfnamefont {N.}~\bibnamefont {Mao}}, \bibinfo {author} {\bibfnamefont {R.}~\bibnamefont {Li}}, \bibinfo {author} {\bibfnamefont {Y.}~\bibnamefont {Dai}}, \bibinfo {author} {\bibfnamefont {B.}~\bibnamefont {Huang}},\ and\ \bibinfo {author} {\bibfnamefont {C.}~\bibnamefont {Niu}},\ }\bibfield  {title} {\bibinfo {title} {Doubled quantum spin hall effect with high-spin chern number in $\ensuremath{\alpha}$-antimonene and $\ensuremath{\alpha}$-bismuthene},\ }\href {https://doi.org/10.1103/PhysRevB.105.195142} {\bibfield  {journal} {\bibinfo  {journal} {Phys. Rev. B}\ }\textbf {\bibinfo {volume} {105}},\ \bibinfo {pages} {195142} (\bibinfo {year} {2022})}\BibitemShut {NoStop}%
\bibitem [{\citenamefont {Lange}\ \emph {et~al.}(2023)\citenamefont {Lange}, \citenamefont {Bouhon},\ and\ \citenamefont {Slager}}]{PhysRevResearch.5.033013}%
  \BibitemOpen
  \bibfield  {author} {\bibinfo {author} {\bibfnamefont {G.~F.}\ \bibnamefont {Lange}}, \bibinfo {author} {\bibfnamefont {A.}~\bibnamefont {Bouhon}},\ and\ \bibinfo {author} {\bibfnamefont {R.-J.}\ \bibnamefont {Slager}},\ }\bibfield  {title} {\bibinfo {title} {Spin texture as a bulk indicator of fragile topology},\ }\href {https://doi.org/10.1103/PhysRevResearch.5.033013} {\bibfield  {journal} {\bibinfo  {journal} {Phys. Rev. Res.}\ }\textbf {\bibinfo {volume} {5}},\ \bibinfo {pages} {033013} (\bibinfo {year} {2023})}\BibitemShut {NoStop}%
\bibitem [{\citenamefont {Peng}\ \emph {et~al.}(2023)\citenamefont {Peng}, \citenamefont {Lange}, \citenamefont {Bennett}, \citenamefont {Slager},\ and\ \citenamefont {Monserrat}}]{Peng_2023}%
  \BibitemOpen
  \bibfield  {author} {\bibinfo {author} {\bibfnamefont {B.}~\bibnamefont {Peng}}, \bibinfo {author} {\bibfnamefont {G.~F.}\ \bibnamefont {Lange}}, \bibinfo {author} {\bibfnamefont {D.}~\bibnamefont {Bennett}}, \bibinfo {author} {\bibfnamefont {R.-J.}\ \bibnamefont {Slager}},\ and\ \bibinfo {author} {\bibfnamefont {B.}~\bibnamefont {Monserrat}},\ }\href@noop {} {\bibinfo {title} {Photo-induced electronic and spin topological phase transitions in monolayer bismuth}} (\bibinfo {year} {2023}),\ \Eprint {https://arxiv.org/abs/2310.16886} {arXiv:2310.16886 [cond-mat.mtrl-sci]} \BibitemShut {NoStop}%
\bibitem [{\citenamefont {Wang}\ \emph {et~al.}(2023)\citenamefont {Wang}, \citenamefont {Hung}, \citenamefont {Zhou}, \citenamefont {Ong},\ and\ \citenamefont {Lin}}]{feature}%
  \BibitemOpen
  \bibfield  {author} {\bibinfo {author} {\bibfnamefont {B.}~\bibnamefont {Wang}}, \bibinfo {author} {\bibfnamefont {Y.-C.}\ \bibnamefont {Hung}}, \bibinfo {author} {\bibfnamefont {X.}~\bibnamefont {Zhou}}, \bibinfo {author} {\bibfnamefont {T.}~\bibnamefont {Ong}},\ and\ \bibinfo {author} {\bibfnamefont {H.}~\bibnamefont {Lin}},\ }\href@noop {} {\bibinfo {title} {Feature spectrum topology}} (\bibinfo {year} {2023}),\ \Eprint {https://arxiv.org/abs/2310.14832} {arXiv:2310.14832 [cond-mat.mtrl-sci]} \BibitemShut {NoStop}%
\bibitem [{\citenamefont {Yao}\ \emph {et~al.}(2023)\citenamefont {Yao}, \citenamefont {Zhou}, \citenamefont {Hung}, \citenamefont {Lin}, \citenamefont {Bansil},\ and\ \citenamefont {Chang}}]{Chang_2024}%
  \BibitemOpen
  \bibfield  {author} {\bibinfo {author} {\bibfnamefont {Y.-T.}\ \bibnamefont {Yao}}, \bibinfo {author} {\bibfnamefont {X.}~\bibnamefont {Zhou}}, \bibinfo {author} {\bibfnamefont {Y.-C.}\ \bibnamefont {Hung}}, \bibinfo {author} {\bibfnamefont {H.}~\bibnamefont {Lin}}, \bibinfo {author} {\bibfnamefont {A.}~\bibnamefont {Bansil}},\ and\ \bibinfo {author} {\bibfnamefont {T.-R.}\ \bibnamefont {Chang}},\ }\href@noop {} {\bibinfo {title} {Feature-energy duality of topological boundary states in multilayer quantum spin hall insulator}} (\bibinfo {year} {2023}),\ \Eprint {https://arxiv.org/abs/2312.11794} {arXiv:2312.11794 [cond-mat.mtrl-sci]} \BibitemShut {NoStop}%
\bibitem [{\citenamefont {Shi}\ \emph {et~al.}(2020{\natexlab{a}})\citenamefont {Shi}, \citenamefont {Li}, \citenamefont {Xue}, \citenamefont {Yuan}, \citenamefont {Lv}, \citenamefont {Xu}, \citenamefont {Jia}, \citenamefont {Gao}, \citenamefont {Chen}, \citenamefont {Zhu},\ and\ \citenamefont {Li}}]{Sb_Li_2020}%
  \BibitemOpen
  \bibfield  {author} {\bibinfo {author} {\bibfnamefont {Z.-Q.}\ \bibnamefont {Shi}}, \bibinfo {author} {\bibfnamefont {H.}~\bibnamefont {Li}}, \bibinfo {author} {\bibfnamefont {C.-L.}\ \bibnamefont {Xue}}, \bibinfo {author} {\bibfnamefont {Q.-Q.}\ \bibnamefont {Yuan}}, \bibinfo {author} {\bibfnamefont {Y.-Y.}\ \bibnamefont {Lv}}, \bibinfo {author} {\bibfnamefont {Y.-J.}\ \bibnamefont {Xu}}, \bibinfo {author} {\bibfnamefont {Z.-Y.}\ \bibnamefont {Jia}}, \bibinfo {author} {\bibfnamefont {L.}~\bibnamefont {Gao}}, \bibinfo {author} {\bibfnamefont {Y.}~\bibnamefont {Chen}}, \bibinfo {author} {\bibfnamefont {W.}~\bibnamefont {Zhu}},\ and\ \bibinfo {author} {\bibfnamefont {S.-C.}\ \bibnamefont {Li}},\ }\bibfield  {title} {\bibinfo {title} {Tuning the electronic structure of an $\alpha$-antimonene monolayer through interface engineering},\ }\href {https://doi.org/10.1021/acs.nanolett.0c03704} {\bibfield  {journal} {\bibinfo  {journal} {Nano Letters}\ }\textbf {\bibinfo {volume} {20}},\ \bibinfo {pages} {8408}
  (\bibinfo {year} {2020}{\natexlab{a}})},\ \bibinfo {note} {pMID: 33064495}\BibitemShut {NoStop}%
\bibitem [{\citenamefont {Märkl}\ \emph {et~al.}(2017)\citenamefont {Märkl}, \citenamefont {Kowalczyk}, \citenamefont {Ster}, \citenamefont {Mahajan}, \citenamefont {Pirie}, \citenamefont {Ahmed}, \citenamefont {Bian}, \citenamefont {Wang}, \citenamefont {Chiang},\ and\ \citenamefont {Brown}}]{Märkl_2018}%
  \BibitemOpen
  \bibfield  {author} {\bibinfo {author} {\bibfnamefont {T.}~\bibnamefont {Märkl}}, \bibinfo {author} {\bibfnamefont {P.~J.}\ \bibnamefont {Kowalczyk}}, \bibinfo {author} {\bibfnamefont {M.~L.}\ \bibnamefont {Ster}}, \bibinfo {author} {\bibfnamefont {I.~V.}\ \bibnamefont {Mahajan}}, \bibinfo {author} {\bibfnamefont {H.}~\bibnamefont {Pirie}}, \bibinfo {author} {\bibfnamefont {Z.}~\bibnamefont {Ahmed}}, \bibinfo {author} {\bibfnamefont {G.}~\bibnamefont {Bian}}, \bibinfo {author} {\bibfnamefont {X.}~\bibnamefont {Wang}}, \bibinfo {author} {\bibfnamefont {T.-C.}\ \bibnamefont {Chiang}},\ and\ \bibinfo {author} {\bibfnamefont {S.~A.}\ \bibnamefont {Brown}},\ }\bibfield  {title} {\bibinfo {title} {Engineering multiple topological phases in nanoscale van der waals heterostructures: realisation of -antimonene},\ }\href {https://doi.org/10.1088/2053-1583/aa8d8e} {\bibfield  {journal} {\bibinfo  {journal} {2D Materials}\ }\textbf {\bibinfo {volume} {5}},\ \bibinfo {pages} {011002} (\bibinfo {year}
  {2017})}\BibitemShut {NoStop}%
\bibitem [{\citenamefont {Bian}\ \emph {et~al.}(2009{\natexlab{a}})\citenamefont {Bian}, \citenamefont {Miller},\ and\ \citenamefont {Chiang}}]{Bi_Meta_2009}%
  \BibitemOpen
  \bibfield  {author} {\bibinfo {author} {\bibfnamefont {G.}~\bibnamefont {Bian}}, \bibinfo {author} {\bibfnamefont {T.}~\bibnamefont {Miller}},\ and\ \bibinfo {author} {\bibfnamefont {T.-C.}\ \bibnamefont {Chiang}},\ }\bibfield  {title} {\bibinfo {title} {Electronic structure and surface-mediated metastability of bi films on si(111)-$7\ifmmode\times\else\texttimes\fi{}7$ studied by angle-resolved photoemission spectroscopy},\ }\href {https://doi.org/10.1103/PhysRevB.80.245407} {\bibfield  {journal} {\bibinfo  {journal} {Phys. Rev. B}\ }\textbf {\bibinfo {volume} {80}},\ \bibinfo {pages} {245407} (\bibinfo {year} {2009}{\natexlab{a}})}\BibitemShut {NoStop}%
\bibitem [{\citenamefont {Lu}\ \emph {et~al.}(2016)\citenamefont {Lu}, \citenamefont {Zhou}, \citenamefont {Chang}, \citenamefont {Guan}, \citenamefont {Chen}, \citenamefont {Jiang}, \citenamefont {Jiang}, \citenamefont {Wang}, \citenamefont {Yang}, \citenamefont {Feng}, \citenamefont {Kawazoe},\ and\ \citenamefont {Lin}}]{Lu2016}%
  \BibitemOpen
  \bibfield  {author} {\bibinfo {author} {\bibfnamefont {Y.}~\bibnamefont {Lu}}, \bibinfo {author} {\bibfnamefont {D.}~\bibnamefont {Zhou}}, \bibinfo {author} {\bibfnamefont {G.}~\bibnamefont {Chang}}, \bibinfo {author} {\bibfnamefont {S.}~\bibnamefont {Guan}}, \bibinfo {author} {\bibfnamefont {W.}~\bibnamefont {Chen}}, \bibinfo {author} {\bibfnamefont {Y.}~\bibnamefont {Jiang}}, \bibinfo {author} {\bibfnamefont {J.}~\bibnamefont {Jiang}}, \bibinfo {author} {\bibfnamefont {X.-s.}\ \bibnamefont {Wang}}, \bibinfo {author} {\bibfnamefont {S.~A.}\ \bibnamefont {Yang}}, \bibinfo {author} {\bibfnamefont {Y.~P.}\ \bibnamefont {Feng}}, \bibinfo {author} {\bibfnamefont {Y.}~\bibnamefont {Kawazoe}},\ and\ \bibinfo {author} {\bibfnamefont {H.}~\bibnamefont {Lin}},\ }\bibfield  {title} {\bibinfo {title} {Multiple unpinned dirac points in group-va single-layers with phosphorene structure},\ }\href {https://doi.org/10.1038/npjcompumats.2016.11} {\bibfield  {journal} {\bibinfo  {journal} {npj Computational Materials}\
  }\textbf {\bibinfo {volume} {2}},\ \bibinfo {pages} {16011} (\bibinfo {year} {2016})}\BibitemShut {NoStop}%
\bibitem [{\citenamefont {Lu}\ \emph {et~al.}(2022)\citenamefont {Lu}, \citenamefont {Cook}, \citenamefont {Zhang}, \citenamefont {Chen}, \citenamefont {Snyder}, \citenamefont {Nguyen}, \citenamefont {Reddy}, \citenamefont {Qin}, \citenamefont {Zhan}, \citenamefont {Zhao}, \citenamefont {Kowalczyk}, \citenamefont {Brown}, \citenamefont {Chiang}, \citenamefont {Yang}, \citenamefont {Chang},\ and\ \citenamefont {Bian}}]{Lu_2022}%
  \BibitemOpen
  \bibfield  {author} {\bibinfo {author} {\bibfnamefont {Q.}~\bibnamefont {Lu}}, \bibinfo {author} {\bibfnamefont {J.}~\bibnamefont {Cook}}, \bibinfo {author} {\bibfnamefont {X.}~\bibnamefont {Zhang}}, \bibinfo {author} {\bibfnamefont {K.~Y.}\ \bibnamefont {Chen}}, \bibinfo {author} {\bibfnamefont {M.}~\bibnamefont {Snyder}}, \bibinfo {author} {\bibfnamefont {D.~T.}\ \bibnamefont {Nguyen}}, \bibinfo {author} {\bibfnamefont {P.~V.~S.}\ \bibnamefont {Reddy}}, \bibinfo {author} {\bibfnamefont {B.}~\bibnamefont {Qin}}, \bibinfo {author} {\bibfnamefont {S.}~\bibnamefont {Zhan}}, \bibinfo {author} {\bibfnamefont {L.-D.}\ \bibnamefont {Zhao}}, \bibinfo {author} {\bibfnamefont {P.~J.}\ \bibnamefont {Kowalczyk}}, \bibinfo {author} {\bibfnamefont {S.~A.}\ \bibnamefont {Brown}}, \bibinfo {author} {\bibfnamefont {T.-C.}\ \bibnamefont {Chiang}}, \bibinfo {author} {\bibfnamefont {S.~A.}\ \bibnamefont {Yang}}, \bibinfo {author} {\bibfnamefont {T.-R.}\ \bibnamefont {Chang}},\ and\ \bibinfo {author} {\bibfnamefont
  {G.}~\bibnamefont {Bian}},\ }\bibfield  {title} {\bibinfo {title} {Realization of unpinned two-dimensional dirac states in antimony atomic layers},\ }\href {https://doi.org/10.1038/s41467-022-32327-8} {\bibfield  {journal} {\bibinfo  {journal} {Nature Communications}\ }\textbf {\bibinfo {volume} {13}},\ \bibinfo {pages} {4603} (\bibinfo {year} {2022})}\BibitemShut {NoStop}%
\bibitem [{\citenamefont {Bian}\ \emph {et~al.}(2013{\natexlab{a}})\citenamefont {Bian}, \citenamefont {Wang}, \citenamefont {Miller},\ and\ \citenamefont {Chiang}}]{Rashba_2013}%
  \BibitemOpen
  \bibfield  {author} {\bibinfo {author} {\bibfnamefont {G.}~\bibnamefont {Bian}}, \bibinfo {author} {\bibfnamefont {X.}~\bibnamefont {Wang}}, \bibinfo {author} {\bibfnamefont {T.}~\bibnamefont {Miller}},\ and\ \bibinfo {author} {\bibfnamefont {T.-C.}\ \bibnamefont {Chiang}},\ }\bibfield  {title} {\bibinfo {title} {Origin of giant rashba spin splitting in bi/ag surface alloys},\ }\href {https://doi.org/10.1103/PhysRevB.88.085427} {\bibfield  {journal} {\bibinfo  {journal} {Phys. Rev. B}\ }\textbf {\bibinfo {volume} {88}},\ \bibinfo {pages} {085427} (\bibinfo {year} {2013}{\natexlab{a}})}\BibitemShut {NoStop}%
\bibitem [{\citenamefont {Bian}\ \emph {et~al.}(2013{\natexlab{b}})\citenamefont {Bian}, \citenamefont {Wang}, \citenamefont {Miller},\ and\ \citenamefont {Chiang}}]{PhysRevB.88.085427}%
  \BibitemOpen
  \bibfield  {author} {\bibinfo {author} {\bibfnamefont {G.}~\bibnamefont {Bian}}, \bibinfo {author} {\bibfnamefont {X.}~\bibnamefont {Wang}}, \bibinfo {author} {\bibfnamefont {T.}~\bibnamefont {Miller}},\ and\ \bibinfo {author} {\bibfnamefont {T.-C.}\ \bibnamefont {Chiang}},\ }\bibfield  {title} {\bibinfo {title} {Origin of giant rashba spin splitting in bi/ag surface alloys},\ }\href {https://doi.org/10.1103/PhysRevB.88.085427} {\bibfield  {journal} {\bibinfo  {journal} {Phys. Rev. B}\ }\textbf {\bibinfo {volume} {88}},\ \bibinfo {pages} {085427} (\bibinfo {year} {2013}{\natexlab{b}})}\BibitemShut {NoStop}%
\bibitem [{\citenamefont {Ahn}\ and\ \citenamefont {Yang}(2017)}]{Ahn_2017}%
  \BibitemOpen
  \bibfield  {author} {\bibinfo {author} {\bibfnamefont {J.}~\bibnamefont {Ahn}}\ and\ \bibinfo {author} {\bibfnamefont {B.-J.}\ \bibnamefont {Yang}},\ }\bibfield  {title} {\bibinfo {title} {Unconventional topological phase transition in two-dimensional systems with space-time inversion symmetry},\ }\href {https://doi.org/10.1103/PhysRevLett.118.156401} {\bibfield  {journal} {\bibinfo  {journal} {Phys. Rev. Lett.}\ }\textbf {\bibinfo {volume} {118}},\ \bibinfo {pages} {156401} (\bibinfo {year} {2017})}\BibitemShut {NoStop}%
\bibitem [{\citenamefont {Shi}\ \emph {et~al.}(2020{\natexlab{b}})\citenamefont {Shi}, \citenamefont {Li}, \citenamefont {Xue}, \citenamefont {Yuan}, \citenamefont {Lv}, \citenamefont {Xu}, \citenamefont {Jia}, \citenamefont {Gao}, \citenamefont {Chen}, \citenamefont {Zhu},\ and\ \citenamefont {Li}}]{Shi_2020}%
  \BibitemOpen
  \bibfield  {author} {\bibinfo {author} {\bibfnamefont {Z.-Q.}\ \bibnamefont {Shi}}, \bibinfo {author} {\bibfnamefont {H.}~\bibnamefont {Li}}, \bibinfo {author} {\bibfnamefont {C.-L.}\ \bibnamefont {Xue}}, \bibinfo {author} {\bibfnamefont {Q.-Q.}\ \bibnamefont {Yuan}}, \bibinfo {author} {\bibfnamefont {Y.-Y.}\ \bibnamefont {Lv}}, \bibinfo {author} {\bibfnamefont {Y.-J.}\ \bibnamefont {Xu}}, \bibinfo {author} {\bibfnamefont {Z.-Y.}\ \bibnamefont {Jia}}, \bibinfo {author} {\bibfnamefont {L.}~\bibnamefont {Gao}}, \bibinfo {author} {\bibfnamefont {Y.}~\bibnamefont {Chen}}, \bibinfo {author} {\bibfnamefont {W.}~\bibnamefont {Zhu}},\ and\ \bibinfo {author} {\bibfnamefont {S.-C.}\ \bibnamefont {Li}},\ }\bibfield  {title} {\bibinfo {title} {Tuning the electronic structure of an $\alpha$-antimonene monolayer through interface engineering},\ }\href {https://doi.org/10.1021/acs.nanolett.0c03704} {\bibfield  {journal} {\bibinfo  {journal} {Nano Letters}\ }\textbf {\bibinfo {volume} {20}},\ \bibinfo {pages} {8408}
  (\bibinfo {year} {2020}{\natexlab{b}})},\ \bibinfo {note} {pMID: 33064495},\ \Eprint {https://arxiv.org/abs/https://doi.org/10.1021/acs.nanolett.0c03704} {https://doi.org/10.1021/acs.nanolett.0c03704} \BibitemShut {NoStop}%
\bibitem [{\citenamefont {Salehitaleghani}\ \emph {et~al.}(2023)\citenamefont {Salehitaleghani}, \citenamefont {Maerkl}, \citenamefont {Kowalczyk}, \citenamefont {Wang}, \citenamefont {Bian}, \citenamefont {Chiang},\ and\ \citenamefont {Brown}}]{Salehitaleghani_2023}%
  \BibitemOpen
  \bibfield  {author} {\bibinfo {author} {\bibfnamefont {S.}~\bibnamefont {Salehitaleghani}}, \bibinfo {author} {\bibfnamefont {T.}~\bibnamefont {Maerkl}}, \bibinfo {author} {\bibfnamefont {P.~J.}\ \bibnamefont {Kowalczyk}}, \bibinfo {author} {\bibfnamefont {X.}~\bibnamefont {Wang}}, \bibinfo {author} {\bibfnamefont {G.}~\bibnamefont {Bian}}, \bibinfo {author} {\bibfnamefont {T.-C.}\ \bibnamefont {Chiang}},\ and\ \bibinfo {author} {\bibfnamefont {S.~A.}\ \bibnamefont {Brown}},\ }\bibfield  {title} {\bibinfo {title} {Moiré pattern modulated topological phase and in-gap edge modes in $\alpha$-antimonene},\ }\href {https://doi.org/https://doi.org/10.1016/j.apsusc.2023.157674} {\bibfield  {journal} {\bibinfo  {journal} {Applied Surface Science}\ }\textbf {\bibinfo {volume} {635}},\ \bibinfo {pages} {157674} (\bibinfo {year} {2023})}\BibitemShut {NoStop}%
\bibitem [{\citenamefont {Shi}\ \emph {et~al.}(2019)\citenamefont {Shi}, \citenamefont {Li}, \citenamefont {Yuan}, \citenamefont {Song}, \citenamefont {Lv}, \citenamefont {Shi}, \citenamefont {Jia}, \citenamefont {Gao}, \citenamefont {Chen}, \citenamefont {Zhu},\ and\ \citenamefont {Li}}]{Shi_2019}%
  \BibitemOpen
  \bibfield  {author} {\bibinfo {author} {\bibfnamefont {Z.-Q.}\ \bibnamefont {Shi}}, \bibinfo {author} {\bibfnamefont {H.}~\bibnamefont {Li}}, \bibinfo {author} {\bibfnamefont {Q.-Q.}\ \bibnamefont {Yuan}}, \bibinfo {author} {\bibfnamefont {Y.-H.}\ \bibnamefont {Song}}, \bibinfo {author} {\bibfnamefont {Y.-Y.}\ \bibnamefont {Lv}}, \bibinfo {author} {\bibfnamefont {W.}~\bibnamefont {Shi}}, \bibinfo {author} {\bibfnamefont {Z.-Y.}\ \bibnamefont {Jia}}, \bibinfo {author} {\bibfnamefont {L.}~\bibnamefont {Gao}}, \bibinfo {author} {\bibfnamefont {Y.-B.}\ \bibnamefont {Chen}}, \bibinfo {author} {\bibfnamefont {W.}~\bibnamefont {Zhu}},\ and\ \bibinfo {author} {\bibfnamefont {S.-C.}\ \bibnamefont {Li}},\ }\bibfield  {title} {\bibinfo {title} {Van der waals heteroepitaxial growth of monolayer sb in a puckered honeycomb structure},\ }\href {https://doi.org/https://doi.org/10.1002/adma.201806130} {\bibfield  {journal} {\bibinfo  {journal} {Advanced Materials}\ }\textbf {\bibinfo {volume} {31}},\ \bibinfo {pages}
  {1806130} (\bibinfo {year} {2019})},\ \Eprint {https://arxiv.org/abs/https://onlinelibrary.wiley.com/doi/pdf/10.1002/adma.201806130} {https://onlinelibrary.wiley.com/doi/pdf/10.1002/adma.201806130} \BibitemShut {NoStop}%
\bibitem [{\citenamefont {Li}\ \emph {et~al.}(2022)\citenamefont {Li}, \citenamefont {Yu}, \citenamefont {Zhang}, \citenamefont {Li}, \citenamefont {Qiao}, \citenamefont {Peng}, \citenamefont {Dong}, \citenamefont {Wang}, \citenamefont {Ma}, \citenamefont {Xiao},\ and\ \citenamefont {Yao}}]{Li_2022}%
  \BibitemOpen
  \bibfield  {author} {\bibinfo {author} {\bibfnamefont {J.}~\bibnamefont {Li}}, \bibinfo {author} {\bibfnamefont {K.}~\bibnamefont {Yu}}, \bibinfo {author} {\bibfnamefont {X.}~\bibnamefont {Zhang}}, \bibinfo {author} {\bibfnamefont {Y.}~\bibnamefont {Li}}, \bibinfo {author} {\bibfnamefont {L.}~\bibnamefont {Qiao}}, \bibinfo {author} {\bibfnamefont {X.}~\bibnamefont {Peng}}, \bibinfo {author} {\bibfnamefont {X.}~\bibnamefont {Dong}}, \bibinfo {author} {\bibfnamefont {Z.}~\bibnamefont {Wang}}, \bibinfo {author} {\bibfnamefont {J.}~\bibnamefont {Ma}}, \bibinfo {author} {\bibfnamefont {W.}~\bibnamefont {Xiao}},\ and\ \bibinfo {author} {\bibfnamefont {Y.}~\bibnamefont {Yao}},\ }\bibfield  {title} {\bibinfo {title} {Controllable growth of $\alpha$- and $\beta$-antimonene by interfacial strain},\ }\href {https://doi.org/10.1021/acs.jpcc.1c10177} {\bibfield  {journal} {\bibinfo  {journal} {The Journal of Physical Chemistry C}\ }\textbf {\bibinfo {volume} {126}},\ \bibinfo {pages} {5022} (\bibinfo {year} {2022})},\
  \Eprint {https://arxiv.org/abs/https://doi.org/10.1021/acs.jpcc.1c10177} {https://doi.org/10.1021/acs.jpcc.1c10177} \BibitemShut {NoStop}%
\bibitem [{\citenamefont {Niu}\ \emph {et~al.}(2021)\citenamefont {Niu}, \citenamefont {Wang}, \citenamefont {Yuan}, \citenamefont {Xue}, \citenamefont {Dou}, \citenamefont {Lv}, \citenamefont {Chen},\ and\ \citenamefont {Li}}]{Niu_2021}%
  \BibitemOpen
  \bibfield  {author} {\bibinfo {author} {\bibfnamefont {Y.-Y.}\ \bibnamefont {Niu}}, \bibinfo {author} {\bibfnamefont {C.-R.}\ \bibnamefont {Wang}}, \bibinfo {author} {\bibfnamefont {Q.-Q.}\ \bibnamefont {Yuan}}, \bibinfo {author} {\bibfnamefont {C.-L.}\ \bibnamefont {Xue}}, \bibinfo {author} {\bibfnamefont {L.-G.}\ \bibnamefont {Dou}}, \bibinfo {author} {\bibfnamefont {Y.-Y.}\ \bibnamefont {Lv}}, \bibinfo {author} {\bibfnamefont {Y.}~\bibnamefont {Chen}},\ and\ \bibinfo {author} {\bibfnamefont {S.-C.}\ \bibnamefont {Li}},\ }\bibfield  {title} {\bibinfo {title} {{Surface step edge-assisted monolayer epitaxy of $\alpha$-antimonene on SnSe2 substrate}},\ }\href {https://doi.org/10.1063/5.0061987} {\bibfield  {journal} {\bibinfo  {journal} {AIP Advances}\ }\textbf {\bibinfo {volume} {11}},\ \bibinfo {pages} {095014} (\bibinfo {year} {2021})},\ \Eprint {https://arxiv.org/abs/https://pubs.aip.org/aip/adv/article-pdf/doi/10.1063/5.0061987/12828722/095014\_1\_online.pdf}
  {https://pubs.aip.org/aip/adv/article-pdf/doi/10.1063/5.0061987/12828722/095014\_1\_online.pdf} \BibitemShut {NoStop}%
\bibitem [{\citenamefont {Lu}\ \emph {et~al.}(2021)\citenamefont {Lu}, \citenamefont {Chen}, \citenamefont {Snyder}, \citenamefont {Cook}, \citenamefont {Nguyen}, \citenamefont {Reddy}, \citenamefont {Chang}, \citenamefont {Yang},\ and\ \citenamefont {Bian}}]{Lu_2021}%
  \BibitemOpen
  \bibfield  {author} {\bibinfo {author} {\bibfnamefont {Q.}~\bibnamefont {Lu}}, \bibinfo {author} {\bibfnamefont {K.~Y.}\ \bibnamefont {Chen}}, \bibinfo {author} {\bibfnamefont {M.}~\bibnamefont {Snyder}}, \bibinfo {author} {\bibfnamefont {J.}~\bibnamefont {Cook}}, \bibinfo {author} {\bibfnamefont {D.~T.}\ \bibnamefont {Nguyen}}, \bibinfo {author} {\bibfnamefont {P.~V.~S.}\ \bibnamefont {Reddy}}, \bibinfo {author} {\bibfnamefont {T.-R.}\ \bibnamefont {Chang}}, \bibinfo {author} {\bibfnamefont {S.~A.}\ \bibnamefont {Yang}},\ and\ \bibinfo {author} {\bibfnamefont {G.}~\bibnamefont {Bian}},\ }\bibfield  {title} {\bibinfo {title} {Observation of symmetry-protected dirac states in nonsymmorphic $\ensuremath{\alpha}$-antimonene},\ }\href {https://doi.org/10.1103/PhysRevB.104.L201105} {\bibfield  {journal} {\bibinfo  {journal} {Phys. Rev. B}\ }\textbf {\bibinfo {volume} {104}},\ \bibinfo {pages} {L201105} (\bibinfo {year} {2021})}\BibitemShut {NoStop}%
\bibitem [{\citenamefont {Datye}\ \emph {et~al.}(2022)\citenamefont {Datye}, \citenamefont {Daus}, \citenamefont {Grady}, \citenamefont {Brenner}, \citenamefont {Vaziri},\ and\ \citenamefont {Pop}}]{Datye_2022}%
  \BibitemOpen
  \bibfield  {author} {\bibinfo {author} {\bibfnamefont {I.~M.}\ \bibnamefont {Datye}}, \bibinfo {author} {\bibfnamefont {A.}~\bibnamefont {Daus}}, \bibinfo {author} {\bibfnamefont {R.~W.}\ \bibnamefont {Grady}}, \bibinfo {author} {\bibfnamefont {K.}~\bibnamefont {Brenner}}, \bibinfo {author} {\bibfnamefont {S.}~\bibnamefont {Vaziri}},\ and\ \bibinfo {author} {\bibfnamefont {E.}~\bibnamefont {Pop}},\ }\bibfield  {title} {\bibinfo {title} {Strain-enhanced mobility of monolayer mos2},\ }\href {https://doi.org/10.1021/acs.nanolett.2c01707} {\bibfield  {journal} {\bibinfo  {journal} {Nano Letters}\ }\textbf {\bibinfo {volume} {22}},\ \bibinfo {pages} {8052} (\bibinfo {year} {2022})},\ \bibinfo {note} {pMID: 36198070},\ \Eprint {https://arxiv.org/abs/https://doi.org/10.1021/acs.nanolett.2c01707} {https://doi.org/10.1021/acs.nanolett.2c01707} \BibitemShut {NoStop}%
\bibitem [{\citenamefont {Bian}\ \emph {et~al.}(2009{\natexlab{b}})\citenamefont {Bian}, \citenamefont {Miller},\ and\ \citenamefont {Chiang}}]{PhysRevB.80.245407}%
  \BibitemOpen
  \bibfield  {author} {\bibinfo {author} {\bibfnamefont {G.}~\bibnamefont {Bian}}, \bibinfo {author} {\bibfnamefont {T.}~\bibnamefont {Miller}},\ and\ \bibinfo {author} {\bibfnamefont {T.-C.}\ \bibnamefont {Chiang}},\ }\bibfield  {title} {\bibinfo {title} {Electronic structure and surface-mediated metastability of bi films on si(111)-$7\ifmmode\times\else\texttimes\fi{}7$ studied by angle-resolved photoemission spectroscopy},\ }\href {https://doi.org/10.1103/PhysRevB.80.245407} {\bibfield  {journal} {\bibinfo  {journal} {Phys. Rev. B}\ }\textbf {\bibinfo {volume} {80}},\ \bibinfo {pages} {245407} (\bibinfo {year} {2009}{\natexlab{b}})}\BibitemShut {NoStop}%
\bibitem [{\citenamefont {Zhang}\ \emph {et~al.}(2021)\citenamefont {Zhang}, \citenamefont {Wang}, \citenamefont {Yang}, \citenamefont {Zhang}, \citenamefont {Xu},\ and\ \citenamefont {Liu}}]{Zhang_2021}%
  \BibitemOpen
  \bibfield  {author} {\bibinfo {author} {\bibfnamefont {H.}~\bibnamefont {Zhang}}, \bibinfo {author} {\bibfnamefont {Y.}~\bibnamefont {Wang}}, \bibinfo {author} {\bibfnamefont {W.}~\bibnamefont {Yang}}, \bibinfo {author} {\bibfnamefont {J.}~\bibnamefont {Zhang}}, \bibinfo {author} {\bibfnamefont {X.}~\bibnamefont {Xu}},\ and\ \bibinfo {author} {\bibfnamefont {F.}~\bibnamefont {Liu}},\ }\bibfield  {title} {\bibinfo {title} {Selective substrate-orbital-filtering effect to realize the large-gap quantum spin hall effect},\ }\href {https://doi.org/10.1021/acs.nanolett.1c01765} {\bibfield  {journal} {\bibinfo  {journal} {Nano Letters}\ }\textbf {\bibinfo {volume} {21}},\ \bibinfo {pages} {5828} (\bibinfo {year} {2021})},\ \bibinfo {note} {pMID: 34156241},\ \Eprint {https://arxiv.org/abs/https://doi.org/10.1021/acs.nanolett.1c01765} {https://doi.org/10.1021/acs.nanolett.1c01765} \BibitemShut {NoStop}%
\bibitem [{\citenamefont {Liu}\ \emph {et~al.}(2022)\citenamefont {Liu}, \citenamefont {Bai}, \citenamefont {Wang}, \citenamefont {Song},\ and\ \citenamefont {Liu}}]{Liu_2022}%
  \BibitemOpen
  \bibfield  {author} {\bibinfo {author} {\bibfnamefont {K.}~\bibnamefont {Liu}}, \bibinfo {author} {\bibfnamefont {K.}~\bibnamefont {Bai}}, \bibinfo {author} {\bibfnamefont {J.}~\bibnamefont {Wang}}, \bibinfo {author} {\bibfnamefont {J.}~\bibnamefont {Song}},\ and\ \bibinfo {author} {\bibfnamefont {Y.}~\bibnamefont {Liu}},\ }\bibfield  {title} {\bibinfo {title} {Phase-dependent epitaxy for antimonene growth on silver substrate},\ }\bibfield  {journal} {\bibinfo  {journal} {Frontiers in Physics}\ }\textbf {\bibinfo {volume} {10}},\ \href {https://doi.org/10.3389/fphy.2022.856526} {10.3389/fphy.2022.856526} (\bibinfo {year} {2022})\BibitemShut {NoStop}%
\bibitem [{\citenamefont {Cook}\ \emph {et~al.}(2023)\citenamefont {Cook}, \citenamefont {Halbertal}, \citenamefont {Lu}, \citenamefont {Zhang}, \citenamefont {Conner}, \citenamefont {Watson}, \citenamefont {Snyder}, \citenamefont {Pollard}, \citenamefont {Hor}, \citenamefont {Basov},\ and\ \citenamefont {Bian}}]{Cook_2023}%
  \BibitemOpen
  \bibfield  {author} {\bibinfo {author} {\bibfnamefont {J.}~\bibnamefont {Cook}}, \bibinfo {author} {\bibfnamefont {D.}~\bibnamefont {Halbertal}}, \bibinfo {author} {\bibfnamefont {Q.}~\bibnamefont {Lu}}, \bibinfo {author} {\bibfnamefont {X.}~\bibnamefont {Zhang}}, \bibinfo {author} {\bibfnamefont {C.}~\bibnamefont {Conner}}, \bibinfo {author} {\bibfnamefont {G.}~\bibnamefont {Watson}}, \bibinfo {author} {\bibfnamefont {M.}~\bibnamefont {Snyder}}, \bibinfo {author} {\bibfnamefont {M.}~\bibnamefont {Pollard}}, \bibinfo {author} {\bibfnamefont {Y.~S.}\ \bibnamefont {Hor}}, \bibinfo {author} {\bibfnamefont {D.~N.}\ \bibnamefont {Basov}},\ and\ \bibinfo {author} {\bibfnamefont {G.}~\bibnamefont {Bian}},\ }\bibfield  {title} {\bibinfo {title} {Moiré modulation of lattice strains in pdte2 quantum films},\ }\href {https://doi.org/10.1088/2053-1583/accc9c} {\bibfield  {journal} {\bibinfo  {journal} {2D Materials}\ }\textbf {\bibinfo {volume} {10}},\ \bibinfo {pages} {035005} (\bibinfo {year} {2023})}\BibitemShut
  {NoStop}%
\bibitem [{\citenamefont {Halbertal}\ \emph {et~al.}(2023)\citenamefont {Halbertal}, \citenamefont {Klebl}, \citenamefont {Hsieh}, \citenamefont {Cook}, \citenamefont {Carr}, \citenamefont {Bian}, \citenamefont {Dean}, \citenamefont {Kennes},\ and\ \citenamefont {Basov}}]{Halbertal_2023}%
  \BibitemOpen
  \bibfield  {author} {\bibinfo {author} {\bibfnamefont {D.}~\bibnamefont {Halbertal}}, \bibinfo {author} {\bibfnamefont {L.}~\bibnamefont {Klebl}}, \bibinfo {author} {\bibfnamefont {V.}~\bibnamefont {Hsieh}}, \bibinfo {author} {\bibfnamefont {J.}~\bibnamefont {Cook}}, \bibinfo {author} {\bibfnamefont {S.}~\bibnamefont {Carr}}, \bibinfo {author} {\bibfnamefont {G.}~\bibnamefont {Bian}}, \bibinfo {author} {\bibfnamefont {C.~R.}\ \bibnamefont {Dean}}, \bibinfo {author} {\bibfnamefont {D.~M.}\ \bibnamefont {Kennes}},\ and\ \bibinfo {author} {\bibfnamefont {D.~N.}\ \bibnamefont {Basov}},\ }\bibfield  {title} {\bibinfo {title} {Multilayered atomic relaxation in van der waals heterostructures},\ }\href {https://doi.org/10.1103/PhysRevX.13.011026} {\bibfield  {journal} {\bibinfo  {journal} {Phys. Rev. X}\ }\textbf {\bibinfo {volume} {13}},\ \bibinfo {pages} {011026} (\bibinfo {year} {2023})}\BibitemShut {NoStop}%
\bibitem [{\citenamefont {Deng}\ \emph {et~al.}(2018)\citenamefont {Deng}, \citenamefont {Sumant},\ and\ \citenamefont {Berry}}]{DENG201814}%
  \BibitemOpen
  \bibfield  {author} {\bibinfo {author} {\bibfnamefont {S.}~\bibnamefont {Deng}}, \bibinfo {author} {\bibfnamefont {A.~V.}\ \bibnamefont {Sumant}},\ and\ \bibinfo {author} {\bibfnamefont {V.}~\bibnamefont {Berry}},\ }\bibfield  {title} {\bibinfo {title} {Strain engineering in two-dimensional nanomaterials beyond graphene},\ }\href {https://doi.org/https://doi.org/10.1016/j.nantod.2018.07.001} {\bibfield  {journal} {\bibinfo  {journal} {Nano Today}\ }\textbf {\bibinfo {volume} {22}},\ \bibinfo {pages} {14} (\bibinfo {year} {2018})}\BibitemShut {NoStop}%
\end{thebibliography}%



\newpage
\begin{figure}
\includegraphics[width=1.0\linewidth]{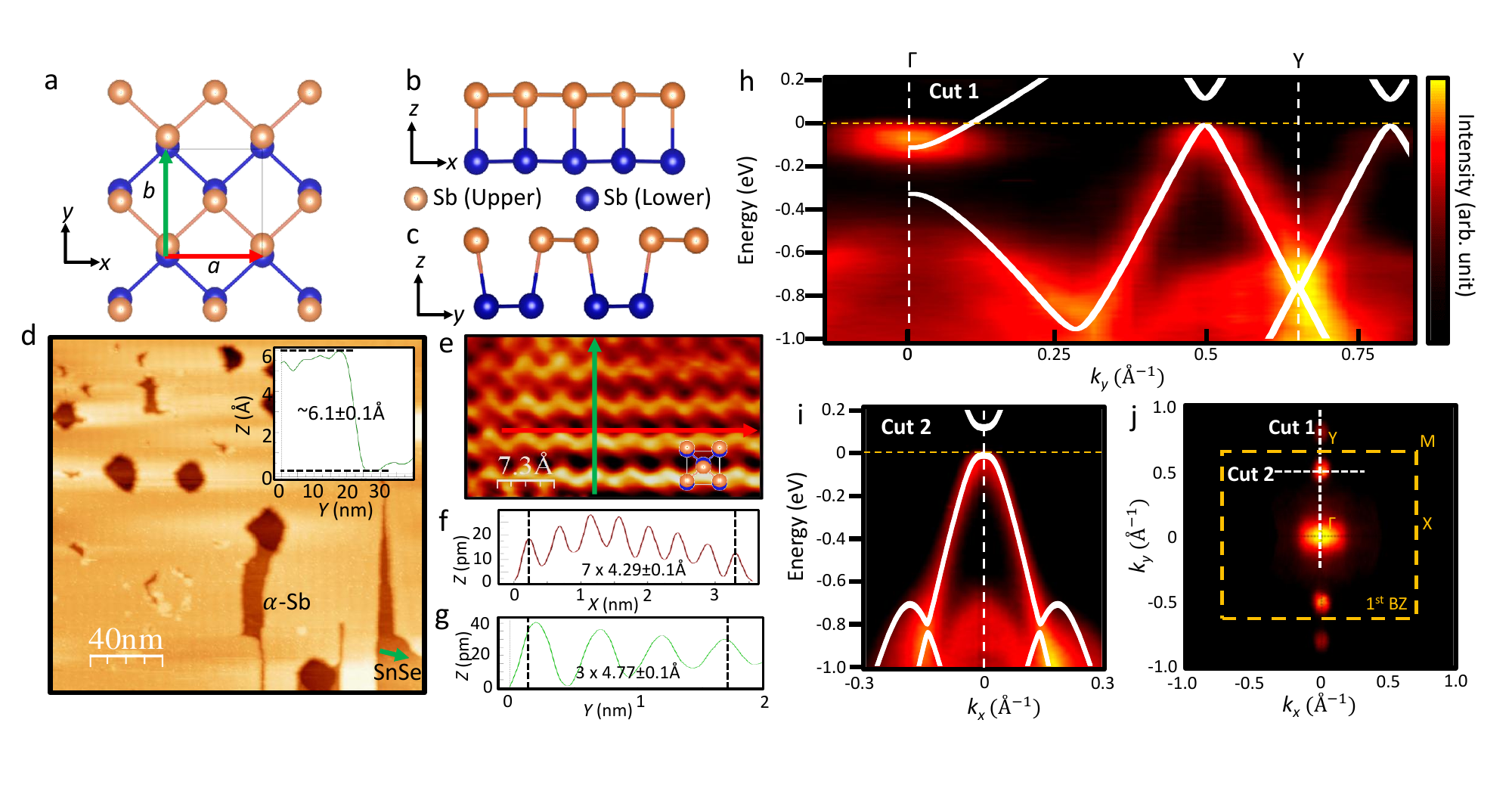}
\caption{(a) Top view of crystal structure of 1ML $\alpha$-Sb. (b,c) Side view of crystal structure of 1ML $\alpha$-Sb along $y$-axis and $x$-axis, respectively. (d) STM image of 1ML $\alpha$-Sb grown on SnSe substrate. The inset shows the height profile taken along the line marked by the green arrow. (e) Atomically resolved STM image of 1ML $\alpha$-Sb on SnSe substrate. The unit cell is marked. (f, g) STM height profile taken along the red ($x$-direction) and green ($y$-direction) lines marked in (e). (h)  ARPES spectrum of 1ML $\alpha$-Sb (grown on SnSe) along $\Gamma-$Y direction. The white solid lines plot the calculated band structure. (i) ARPES spectral cut taken along the line of ``Cut2'' marked in (j). (j) ARPES Fermi surface of 1ML $\alpha$-Sb on SnSe.}%
\end{figure}

\newpage

\begin{figure}
\includegraphics[width=1.0\linewidth]{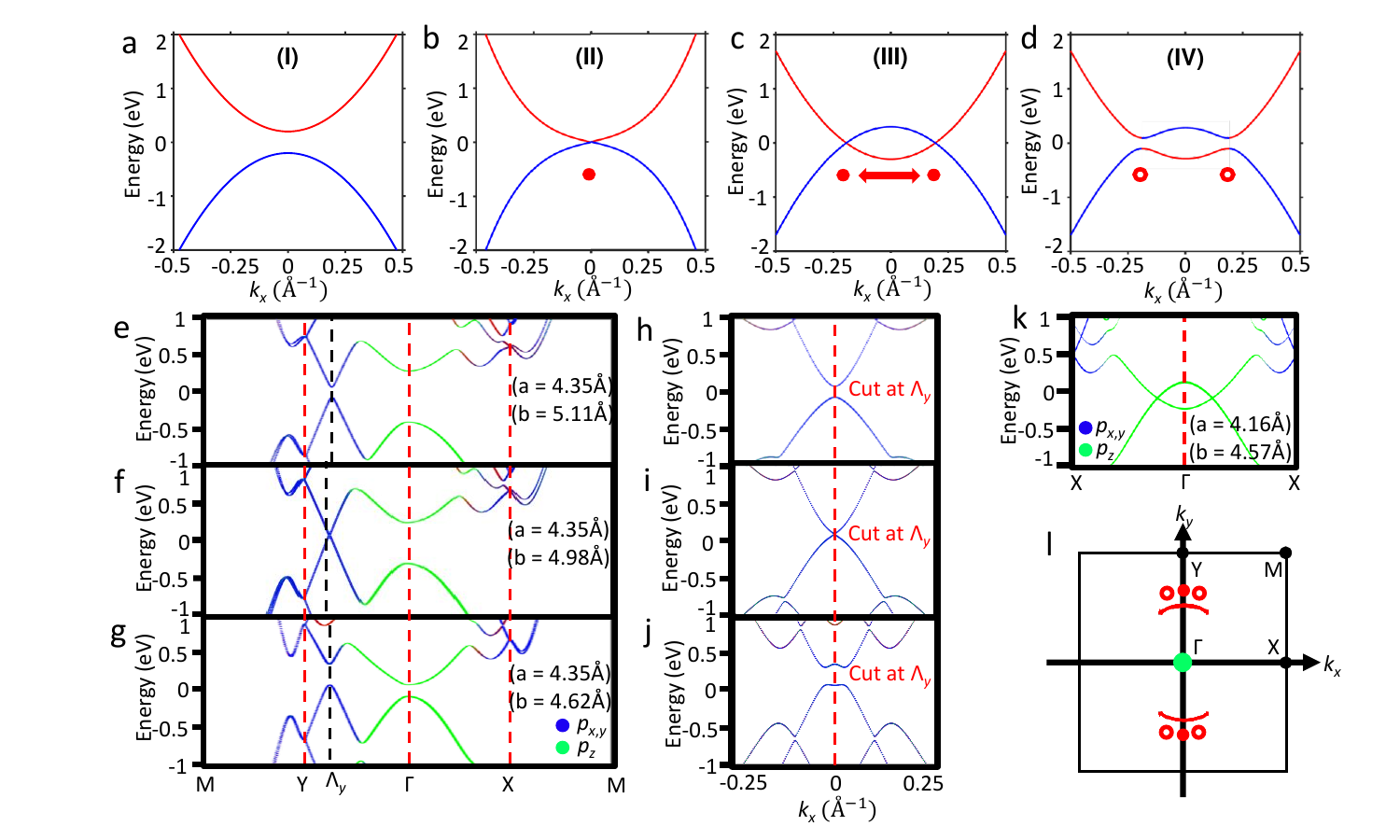}
\caption{ Schematic of band structure: (a) an energy gap separates valence and conduction bands; (b) valence and conduction bands touch and form a Dirac point; (c) band inversion occurs in the absence of SOC, forming two Dirac nodes away from the high symmetry line; (d) SOC opens gaps at Dirac points.  (e-g) Calculated band structure of 1ML $\alpha$-Sb with different lattice parameters. The bands are colored according to the orbital components of the states. (h-j) Calculated band dispersion across the valley point $\Lambda_y$ orthogonal to $\Gamma-$Y with the respective lattice parameters in (e-g). (k) Calculated band structure of 1ML $\alpha$-Sb along X$-\Gamma-$X direction. The lattice constants are $a=4.16\mathrm{\AA}$ and $b=4.57\mathrm{\AA}$. A band inversion occurs around the $\Gamma$ point. (l) The first Brillouin zone of $\alpha$-Sb with TRIM points labeled. The red dots and circles correspond to the location of gapless Dirac points and gapped Dirac points shown in (i) and (j), respectively. The green dot at $\Gamma$ indicate the location of band inversion between the conduction and valence bands of $p_z$ orbital character.} 
\end{figure}

\newpage
\begin{figure}
\includegraphics[width=1.0\linewidth]{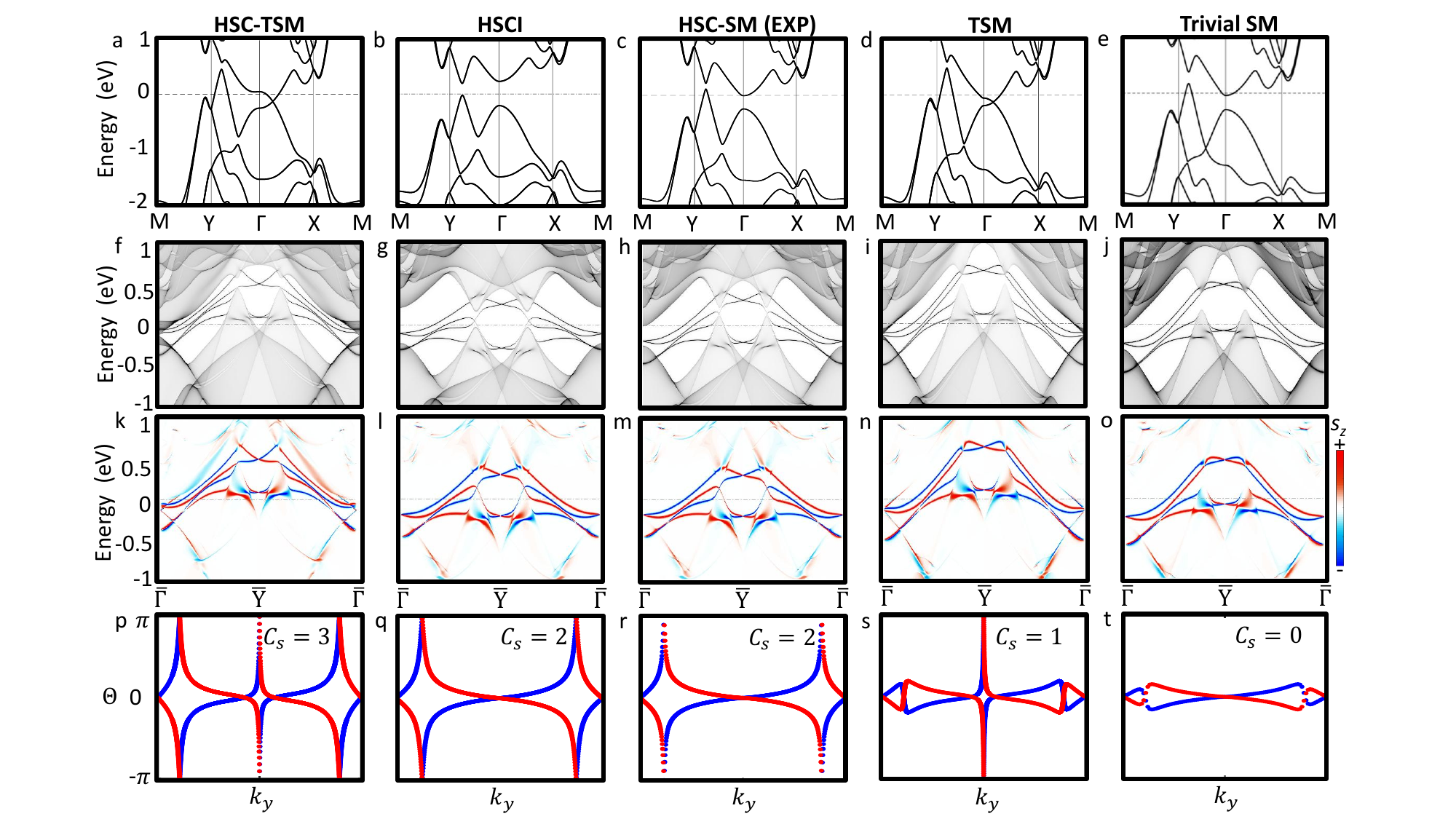}
\caption{Calculated band structure of 1ML $\alpha$-Sb with lattice constants (a) $a=4.16\mathrm{\AA}$ and $b=4.57\mathrm{\AA}$, (b) $a=4.42\mathrm{\AA}$ and $b=4.77\mathrm{\AA}$, (c) $a=4.29\mathrm{\AA}$ and $b=4.77\mathrm{\AA}$ (our experimental value), (d) $a=4.08\mathrm{\AA}$ and $b=4.77\mathrm{\AA}$, and (e) $a=4.16\mathrm{\AA}$ and $b=4.96\mathrm{\AA}$. (f-j) Calculated edge state bands along $\overline{\mathrm{Y}}-\overline{\Gamma}-\overline{\mathrm{Y}}$ direction of a semi-infinite 1ML $\alpha$-Sb slab with lattice constants used in (a-e), respectively. (k-o) Calculated spin polarization $\langle s_z\rangle$
of edge states in (f-j), respectively. (p-t) Calculated spin-resolved Wannier center 
$\mathrm{\theta}$ as a function of $k_y$ for 1ML $\alpha$-Sb in (a-e), respectively.The red (blue) curves are for spin-up (spin-down) projected bulk valence bands.
}
\end{figure}

\newpage
\begin{figure}
\includegraphics[width=1.0\linewidth]{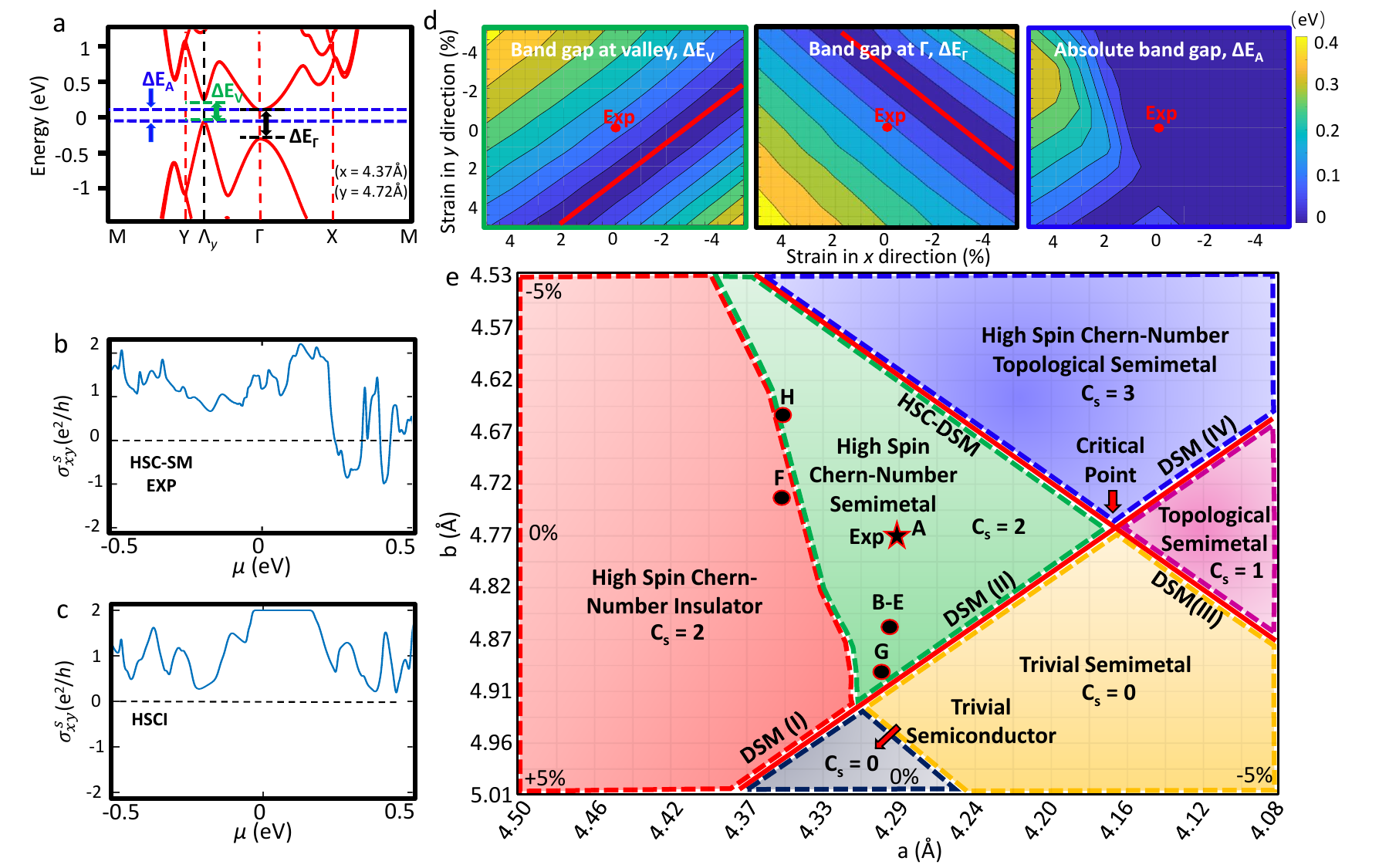}
\caption{(a) Graphical definition of 
energy gaps $\Delta \mathrm{E_A}$, $\Delta \mathrm{E_V}$, and $\Delta \mathrm{E_\Gamma}$. (b) Calculated spin Hall conductivity of $\alpha$-Sb in the high spin Chern-number semimetal phase.
The lattice constants are $a=4.29\mathrm{\AA}$ and $b=4.77\mathrm{\AA}$, as obtained from our experiments. (c) Calculated spin Hall conductivity of $\alpha$-Sb in the high spin Chern-number insulator phase.
The lattice constants are $a=4.42\mathrm{\AA}$ and $b=4.77\mathrm{\AA}$. (d) Maps of energy gaps $\Delta \mathrm{E_V}$, $\Delta \mathrm{E_\Gamma}$, and $\Delta \mathrm{E_A}$ under different lattice strains. (e) Phase diagram of $\alpha$-Sb 
under different lattice strains. The black dots mark lattice constant values extracted from sources in the literature; A\cite{Shi_2020}, B\cite{Salehitaleghani_2023}, C\cite{Märkl_2018}, D\cite{Shi_2019}, E\cite{Li_2022}, F\cite{Bai_2022}, G\cite{Niu_2021}, and H\cite{Lu_2021}.
}%
\end{figure}

\end{document}